\documentclass[prl,nofootinbib,floatfix, 12pt]{revtex4-1}

\usepackage{graphicx}
\usepackage{amsmath}
\usepackage{color}
\usepackage{ulem}

\newcommand{\ket}[1]{|{#1}\rangle}
\newcommand{\bra}[1]{\langle {#1}|}

\newcommand{\be}{\begin{equation}}
\newcommand{\ee}{\end{equation}}

\newcommand{\comment}[1]{}
\definecolor{mypink1}{rgb}{0.858, 0.188, 0.478}
\definecolor{byzantium}{rgb}{0.44, 0.16, 0.39}
\definecolor{mygray}{gray}{0.6}

\begin{document}

\title{Spontaneous Emission in a Matter-Wave Open Quantum System}

\author{Ludwig Krinner, Michael Stewart, Arturo Pazmi\~no, Joonhyuk Kwon, and Dominik Schneble}
\email[Corresponding author: ]{dominik.schneble@stonybrook.edu}

\affiliation{Department of Physics and Astronomy, Stony Brook University, Stony Brook, New York 11794-3800, USA}

\date{\today}

\begin{abstract}
One of the paradigms of a small quantum system in a dissipative environment is the decay of an excited atom undergoing spontaneous photon emission into the fluctuating quantum electrodynamic vacuum. Recent experiments have demonstrated that the gapped photon dispersion in periodic structures can give rise to novel spontaneous-decay behavior including the formation of dissipative bound states. So far, these effects have been restricted to the optical domain. Here, we experimentally demonstrate similar behavior in a system of artificial atoms in an optical lattice that decay by emitting matter-wave, rather than optical, radiation into free space. By controlling the vacuum coupling and excitation energy, we directly observe exponential and partly reversible, non-Markovian dynamics and detect a tunable bound state containing evanescent matter waves for emission at negative excitation energies. Our system provides a flexible platform for the emulation of open-system quantum electrodynamics and studies of dissipative many-body physics with ultracold atoms. 
\end{abstract}

\maketitle

The Weisskopf-Wigner model of spontaneous emission \cite{Weisskopf1930,Milonni1994} is a central concept in quantum optics \cite{Meystre2007}, describing how an excited atom can decay to its ground state due to coupling to zero-point oscillations of the electromagnetic vacuum. It simultaneously represents one of the first open quantum systems discussed in the literature, and an area of research that has recently seen a resurgence of intense theoretical efforts \cite{Rivas2014, Rotter2015, Breuer2016, Vega2017}. In its usual Markovian formulation, the model makes the assumption that the decay proceeds on a much slower time scale than the optical period, which leads to a memoryless, exponential decay of the excited-state amplitude and to an associated Lamb-shift of the transition frequency. For free-space emission, the Markovian approximation is generally fulfilled to high accuracy.

On the other hand, modifications to the mode density of the vacuum can change the features of spontaneous decay. This was first recognized in the 1940s \cite{Purcell1946} and again decades later \cite{Kleppner1981} in the development of cavity quantum electrodynamics \cite{Miller2005,Walther2006,Haroche2006}, where the decay can be altered to the extreme point of coherent vacuum Rabi oscillations when the spectrum is restricted to a single mode.

Between these two limits lies the regime of a vacuum with a bounded continuous spectrum, in which a strong modification of spontaneous decay behavior occurs close to the boundary. An example is given by photonic crystals (also called photonic bandgap materials) \cite{John1987,Yablonovich1987}, where a periodic spatial modulation of the refractive index gives rise to a gapped dispersion relation. For emission close to a bandgap, the Markovian approximation can no longer be made, and novel features appear including oscillatory decay dynamics for energies above the band edge and the formation of atom-photon bound states below \cite{Lambropoulos2000}. Over the past two decades, experiments on spontaneous emission in photonic bandgap materials, including the microwave domain, have observed some of these effects, specifically modified spontaneous emission rates \cite{Tocci1996,Lodahl2004} and Lamb shifts \cite{Liu2010}, as well as spectral signatures for non-exponential decay \cite{Hoeppe2012}. Very recently, experiments have probed the long-predicted bound state \cite{Bykov1975,John1990}, both using transmon qubits coupled to corrugated microwave guides \cite{Liu2017}, and atoms in photonic-crystal waveguides \cite{Hood2016} with the prospect of engineering systems with optical long-range interactions \cite{Douglas2015}.

Here, we realize an atom-optical analog \cite{Vega2008,Navarrete2011,Stewart2017} of emission in a one-dimensional photonic bandgap material, where the singularity in the mode density near the edge of the continuum leads to particularly strong deviations from Markovian behavior. Our system is comprised of elementary matter-wave emitters (occupational spins) in an optical lattice that emit single atoms into free space, with a mechanism described by the Weisskopf-Wigner model. The free tunability of the excitation energy and decay strength allows for a systematic exploration of the emergence of non-Markovian dynamics, including partial reversibility and the formation of a matter-wave bound state which can be directly detected. Importantly, the close spacing of emitters in the optical lattice gives rise to collectively enhanced dynamics beyond the Weisskopf-Wigner model.

\paragraph{\bf Introduction to the system.}

\begin{figure}[ht!]
\centering
    \includegraphics[width=0.5\columnwidth]{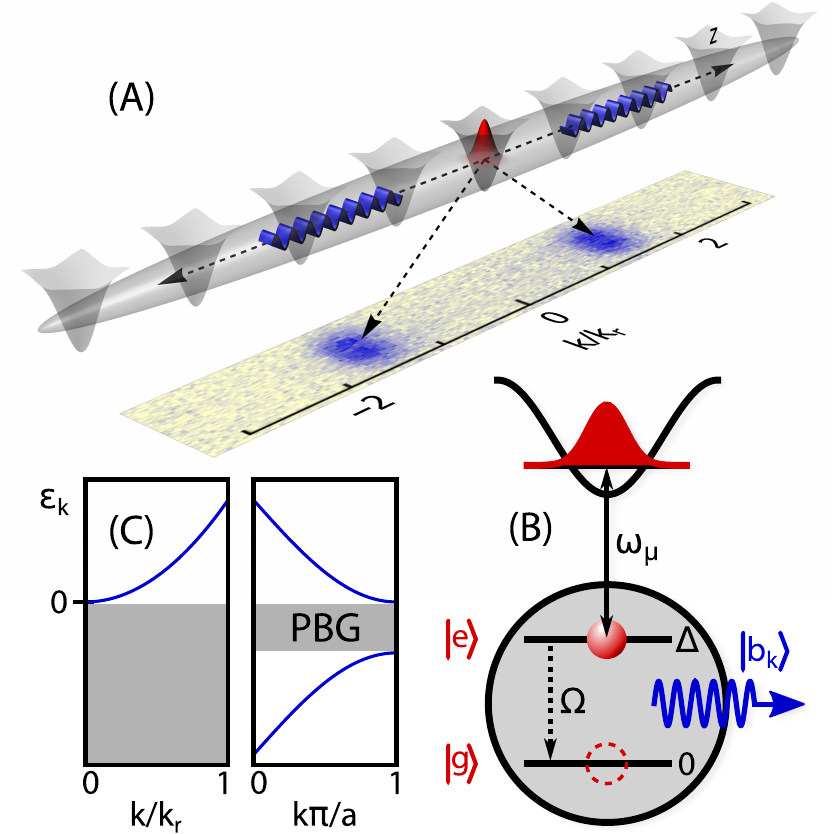}
    \caption{(A) Experimental configuration: An occupied site of an optical lattice embedded in a single-mode matter \mbox{waveguide} acts as an atom emitter; adjacent empty lattice sites act as absorbers. The bottom panel shows a momentum distribution in the waveguide, observed after release and free expansion, where $\hbar k_r=\hbar 2\pi/\lambda_z$ is the recoil momentum. (B) Emission mechanism: an oscillatory coupling field (adjustable frequency $\omega_\mu$, strength $\Omega$) connects the confined atom's internal state $\ket{r}$ (red) to an unconfined state $\ket{b}$ (blue). In a frame rotating at $\omega_{\mu}$, the confined state appears as the excited state $\ket{e}$ of a two-level matter-wave emitter with excitation energy $\hbar\Delta$, whose ground state $\ket{g}$ is the empty site. A transition from $\ket{e}$ to $\ket{g}$ is linked to the appearance of a $\ket{b_k}$ atom at momentum $k$ and energy $\varepsilon_k$. (C) The emission of atoms bears close similarity to the emission of photons in photonic crystals, both featuring quadratic dispersions and energetically forbidden regions.}
\label{FIG:SystemIntro}
\end{figure}

The experimental configuration is shown in Fig.~\ref{FIG:SystemIntro}(A).  Using a deep three-dimensional optical lattice with state-selectivity along one axis, we prepare a sparse array of atoms confined to sites that are embedded in isolated tubes acting as one-dimensional waveguides. An atom's internal state is coherently coupled to a second, unconfined internal state using an oscillatory magnetic field. Each site thus acts as a two-level matter-wave emitter, with harmonic-oscillator ground state occupational levels $\ket{g}$ (empty) and $\ket{e}$ (full), supporting both the emission (for $\ket{e}\to\ket{g}$) or absorption (for $\ket{g}\to\ket{e}$) of an atom. The excitation energy of the emitter, which is given by the detuning $\Delta$ of the coherent coupling from the atomic resonance, is converted into kinetic energy for atomic motion along the axis of the waveguide.

One of the main features of each matter-wave emitter is its ability to undergo spontaneous decay as understood by Weisskopf and Wigner. Assuming no lattice potential, the driven atom performs simple Rabi oscillations between two internal states $\ket{r}$ and $\ket{b}$ described by the Hamiltonian $\hat{H}=(\hbar\Omega/2)e^{-i\delta t} \hat{r} \hat{b}^{\dagger} + \mathrm{H.c}.$, where $\Omega$ denotes the strength and $\delta$ the detuning of the coupling from the transition. The tight confinement of just one of the states (say $\ket{r}$) strongly couples the atom's internal and motional degrees of freedom, producing a zero-point energy shift $\bar{\varepsilon}_0 = \hbar\omega_0/2\gg\hbar\Omega$ in the potential, as well as a kinetic-energy shift $\varepsilon_k = \hbar^2k^2/2m$ for motion of the free $\ket{b}$ state at $\hbar k$ momentum. As a consequence, the detuning and strength of the coupling are shifted to $\Delta_k=\delta+(\bar{\varepsilon}_0-\varepsilon_k)/\hbar$ and $\Omega_k=\Omega\gamma_k$, respectively, with $\gamma_k=\left\langle k |\psi_e\right\rangle$ denoting the overlap of the external wavefunctions. Integration over all possible momenta $k$ then yields  \cite{Stewart2017}
$
\hat{H}=\sum_k~\hbar g_k e^{-i\Delta_k t}\ket{g}\bra{e}\hat{b}^{\dagger}_k+ \rm{H.c}
$ with $g_k=\Omega_k/2$, i.e. the standard Weisskopf-Wigner Hamiltonian describing spontaneous emission into a vacuum of modes ($k,\varepsilon_k$) (see Fig. \ref{FIG:SystemIntro}(B)).

In contrast to optical emission in free space, the dispersion relation $\varepsilon_k$ is quadratic as in a photonic crystal, cf. Fig.~\ref{FIG:SystemIntro}(C). In such crystals, the emission energy relative to the edge of the continuum may be adjusted through the crystal's band structure; in our system, the excitation energy $\hbar\Delta \equiv \hbar\Delta_{k=0}$ itself is tunable including the case $\Delta<0$.  Importantly, the tunability also includes the vacuum coupling $g_k$, which is set by $\Omega$.

\paragraph{\bf Experimental implementation.} In the experiment we use $^{87}$Rb atoms in the hyperfine ground states $\ket{r}=\ket{F=1,m_F=-1}$ and $\ket{b}=\ket{2,0}$ (the fact that $\ket{r}$ lies below $\ket{b}$ is inconsequential in the rotating frame). The atoms are confined to a two-dimensional array of $\sim 10^3$ isolated lattice tubes spaced at $532$~nm, each with a radial confinement of $\omega_{\bot}=2 \pi \times 26~\rm{kHz}$ and a residual axial confinement of $\omega_z=2\pi\times97$~Hz that quantizes the mode spectrum for released $\ket{b}$ atoms in the $z$ direction, but is inconsequential for times much shorter than $\tau_z=2\pi/\omega_z\sim10~\mathrm{ms}$. A state-selective lattice with period $\lambda_z/2=395~\rm{nm}$  and harmonic-oscillator frequency $\omega_0=2\pi\times40(1)~\rm{kHz}$ strongly confines the $\ket{r}$ atoms along the tube axis.

Starting with an optically-trapped Bose-Einstein condensate \cite{Pertot2009}, we first adiabatically create an atomic Mott insulator of $\ket{r}$ atoms by ramping up the lattices, and then transfer a fraction of $\sim 0.82$ to an intermediate state $\ket{2,1}$ for subsequent removal with a short pulse of resonant light. This yields a sample of $2.8(2)\times10^4$ $\ket{r}$ atoms with an average site occupation $\left\langle n_i \right \rangle \lesssim 0.5$ in the tubes. Having thus created an initial state of atomic emitters, we then switch on a 6.8~GHz microwave field of variable coupling strength $\Omega$ and detuning $\Delta$. After a variable time $t$, we measure the population remaining in the lattice and access the momentum distribution of the released atoms with state-selective absorptive imaging, using a combination of band-mapping and Stern-Gerlach separation in time-of-flight (see methods).

\paragraph{\bf Observing spontaneous decay.}

\begin{figure}[h!!]
\centering
    \includegraphics[width=0.5\columnwidth]{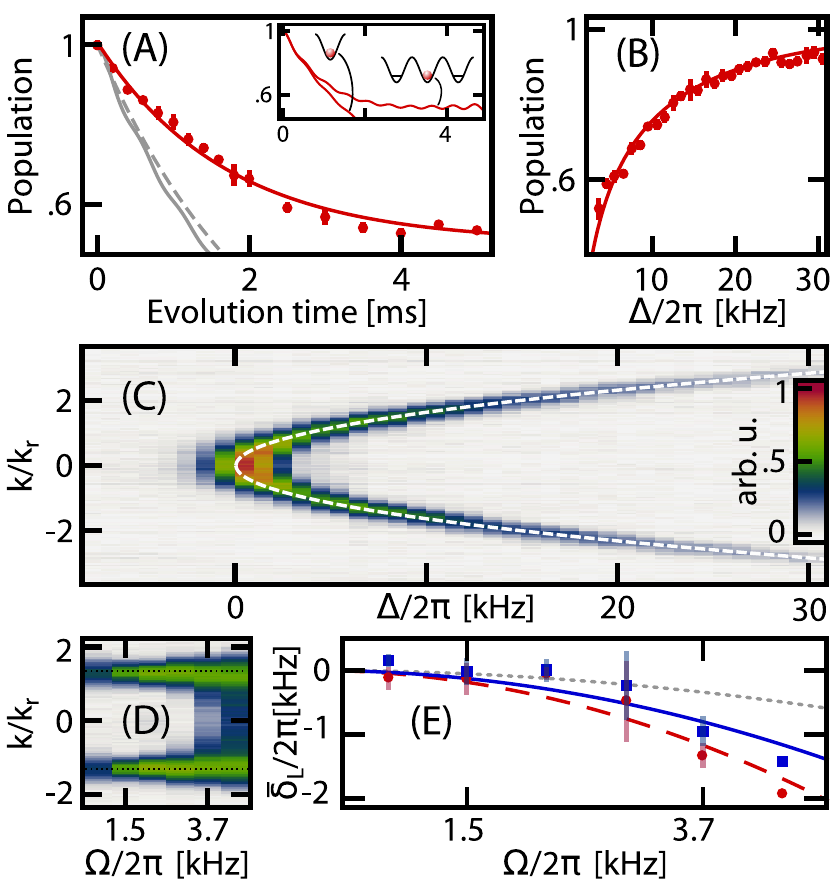}
    \caption{Markov Regime. (A) Time evolution of the lattice population for $\Omega=2\pi\times0.74(5)~$kHz and $\Delta=2\pi\times1.9(3)$ kHz. Each point is the average of at least 3 measurements, bars are the standard error of the mean. The red line is an exponential decay with a rate of $2\pi\times 94(3)$~Hz and an offset of $0.50$. The light gray lines represent the Markov approximation (dashed, $\Gamma=2\pi\times72(12)$~Hz) and the exact analytical solution for an isolated emitter \cite{Stewart2017} (solid). Inset: Simulated decay of a single excited emitter located on one site or in the center of a three-site array (same parameters as in the main figure). (B) Lattice population as a function of $\Delta$ for $t=0.4~$ms and $\Omega=2\pi\times1.5(1)~$kHz. The solid line is the Markovian expectation with the overall decay width $\Gamma$ scaled by 0.61(1). (C) Detected momentum distribution of $\ket{b}$ atoms versus $\Delta$ for parameters as in  (B). The dashed line is the single-particle dispersion. (D)  Raw TOF data for extracting the energy shift at $\Delta=2\pi\times6.0(3)~$kHz. (E) Measured shifts $\bar{\delta}_L=\Delta_{\bar{k}}-\Delta$ in the regime $\Omega/\Delta<1$, for $\tilde{\Omega} t=1.24$ and averaged over $\Delta=2\pi\times 1,2,4,6$~kHz.  The data are extracted from the second moment (maximum) of the momentum distribution, shown using blue squares (red circles). The blue solid and red dashed lines are quadratic fits, the gray dotted line represents $\delta_L(\Omega)$.}
    \label{FIG:Markov}
\end{figure}

A common scenario considered in the Wigner-Weisskopf model is emission deep into the continuum, such that the decay  dynamics is much slower than the time scale set by the excited-state energy (or the elevation above the band edge in the case of a photonic crystal). This allows for a Markov treatment and results in exponential decay of the excited state. Following Fermi's Golden Rule, the decay width $\Gamma$ is the product of the mode density $\rho$ and the square of a matrix element $H_{fi}$ which for optical decay is the product of the electric dipole moment and the zero-point field of the resonant mode. For our system, an analogous analysis \cite{Stewart2017} (valid for $\Omega/\Delta\ll1$) leads to $\Gamma = \Omega^2_{\bar{k}}/\sqrt{\omega_0\Delta}$, containing the 1D mode density $\rho\propto 1/\sqrt{\Delta}$ and $H_{fi} \propto \hbar\Omega_{\bar{k}}$, where $\bar{k}=\sqrt{2m\Delta/\hbar}$ labels the resonant mode.

Because of the residual axial tube confinement, all measurements are taken for $\Omega/\omega_z>1$ and associated time scales shorter than $\tau_z$. The measured $\ket{r}$ population is shown in Fig. \ref{FIG:Markov} for near-Markov parameters as a function of time (A) and detuning (B); the data in (A) are for $\Omega/\Delta\approx0.4$ at $\Delta = 2 \pi\times 2$~kHz (with $\Gamma=2\pi\times72$~Hz). We observe an irreversible, exponential decay in agreement with the expectation; however, the measured population does not decay to zero but instead saturates at a finite value. This discrepancy is readily explained if one takes into account that an excited emitter is not isolated but part of a (mostly) ground-state array that enables reabsorption, in analogy to an optically thick medium. In the Weisskopf-Wigner formalism, the array is modeled by introducing site-dependent phases and projectors, resulting in \cite{Vega2008} $\hat{H}=\sum_{k,j}~\hbar g_k e^{-i(\Delta_k t-k z_j)}\ket{g_i}\bra{e_j}\hat{b}^{\dagger}_k+ \rm{H.c}$. We simulate the dynamics of this model numerically and find that even small arrays reproduce the salient features (inset of Fig. \ref{FIG:Markov} (A); see methods); within $t\sim1/\Gamma=2$~ms, the population gets trapped and only much later escapes from the (hypothetical) edges of the array. At early times, the population $\exp{[-\Gamma(\Delta)t]}$ scales with the detuning $\Delta$ as expected for an isolated emitter (albeit with a sight rescaling of the overall decay rate), as seen in (B).

We next characterize the momentum distribution of the emitted atoms. For this purpose we apply a 0.4~ms long coupling pulse and then observe the location of the $\ket{b}$ atoms after $15$~ms of free fall. Based on the Markov approximation, isotropic emission with wave-packets centered near the resonant momentum $\bar{k}(\Delta)$ is expected. Figure \ref{FIG:Markov} (C) shows the observed momentum distribution as a function of $\Delta$; the emission clearly traces the parabolic dispersion. Moreover, the spectral width decreases, in qualitative agreement with the expectation (while a quantitative comparison is compromised by the finite time of flight). The ``intensity'' of the emitted matter-wave pulse strongly depends on the detuning as already seen in Fig. \ref{FIG:Markov} (B).

The standard Markov treatment of the Weisskopf-Wigner model yields a Lamb shift of the ground and excited states as a unitary coupling to the vacuum. An analogous analysis for our system \cite{Vega2008,Stewart2017} yields a shift $\delta_L = \Omega^2/\omega_0$ of the excited-state energy, to $(\Delta-\delta_L)$.
We measure the momentum distribution for variable $\Omega$ at several values of $\Delta$ and then calculate the mean kinetic energy of the wave-packets both from the second moment of the momentum distributions and from the location of their fitted maxima (see methods for details). To facilitate comparison with the model, the data are taken for a constant effective pulse area $\tilde{\Omega} t$, where $\tilde{\Omega}=(\Omega/\omega_0)^{1/3} \Omega$, and $t\leq1$~ms to mitigate propagation effects. The results for the (quasi-)Markovian regime $\Omega/\Delta<1$ are shown in Fig. \ref{FIG:Markov} (D,E) as a function of $\Omega$. The extracted shift has the sign and approximate quadratic dependence of $\delta_L$, but is a factor of roughly three too large. We caution that, while this alone could point toward the existence of collective enhancement, there is no indication for superradiance \cite{Dicke1954,Vega2008} from the decay data.

\paragraph{\bf Non-Markovian dynamics.}

\begin{figure}[ht!]
\centering
    \includegraphics[width=0.5\columnwidth]{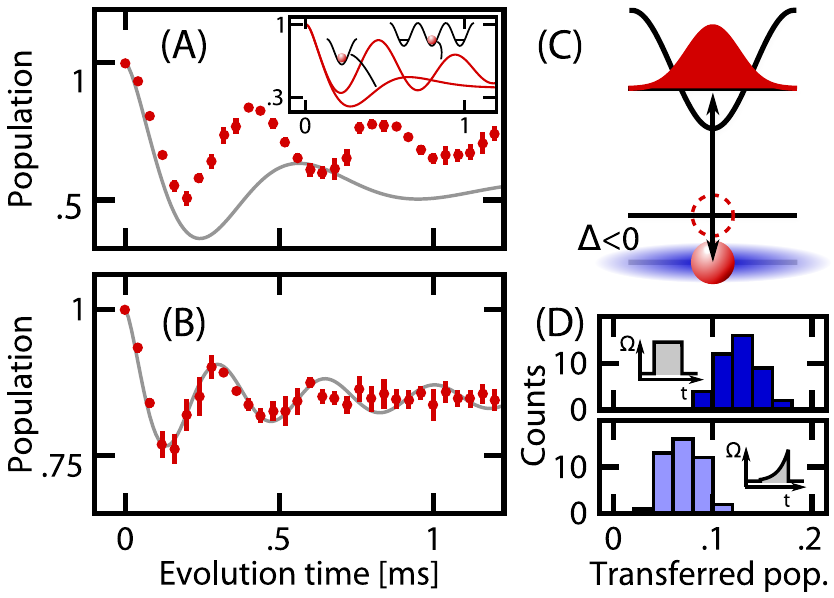}
    \caption{Non-Markovian dynamics and bound-state formation. (A) Time evolution of the $\ket{r}$ population for $\Delta= - 2\pi\times 0.1(3)$~kHz and $\Omega=2\pi\times3.0(3)$~kHz. Each point is the average of at least 3 measurements, with error bars representing the standard error of the mean. The gray line is our analytical model. Inset: Simulated dynamics of a single excited emitter, both isolated, and in a three-site array (same parameters as in main plot). (B) Same as in (A) but for $\Delta= -2\pi\times 1.7(3)$~kHz. The fit parameters of the analytic model are $\Delta= - 2\pi\times 2.08(3)$~kHz and $\Omega=2\pi\times2.79(4)$~kHz.  (C) Illustration of a stationary bound state featuring a localized, evanescent matter wave. (D) Asymptotic fraction of $\ket{b}$ atoms after $t=4$~ms, for $\Omega$ as in (A) and $\Delta=-2\pi\times2.2$~kHz. The top (bottom) panel is for a sudden (adiabatic) turn-on of the coupling. The histograms represent the outcomes of 50 experimental runs each.}
    \label{FIG:NonMarkov}
\end{figure}
Our system readily allows for the study of spontaneous emission outside the Markov regime. In particular, the diverging one-dimensional mode density near $\varepsilon_k=0$ greatly enhances the effects of the edge of the continuum. For emission at low excitation energy, $\Delta/\Omega\ll1$, we expect dynamics reminiscent of a two-level system, with damping provided by low-energy modes. Results of  measurements at $\Delta=0$ are shown in Fig.\ref{FIG:NonMarkov} (A).  We observe oscillations similar to the predictions of our isolated-emitter model \cite{Stewart2017} but with higher frequency and less damping. We numerically integrate our multi-site model $\hat{H}^\prime$ for small arrays and find behavior that much more closely matches to our observations (see inset of Fig. \ref{FIG:NonMarkov} (A)), suggesting that the dynamics are coherently enhanced by low-energy modes, whose wavelengths can extend over several emitters.

\paragraph{\bf Bound-state formation.} For emission below the continuum edge, i.e. for $\Delta<0$, we expect the formation of a stationary bound state \cite{Vega2008,Stewart2017}, as illustrated in Fig.~\ref{FIG:NonMarkov} (C). For our one-dimensional system, this state consists of a partly excited emitter dressed by an evanescent, exponentially decaying matter wave, with a binding energy $\approx\hbar|\Delta|$ and localization length $\xi\approx 1/\sqrt{2 m |\Delta|/\hbar}$ \cite{Stewart2017}. To isolate the bound state, the coupling needs to be turned on slowly to prevent the additional population of freely propagating modes \cite{Stewart2017} representing a nonadiabatic, transient shedding of matter waves. However, we first proceed as before by switching on the coupling, at $\Delta=- 2\pi\times 1.7~$kHz. The lattice population, shown in Fig. \ref{FIG:NonMarkov} (B), shows a transient oscillation settling to an asymptotic value below unity (with an observable $\ket{b}$ population, cf. Fig. 2 (C)). Remarkably, our single-emitter model \cite{Stewart2017} now closely fits the data within the experimental uncertainties. Indeed, for the chosen parameters, $\xi$ is less than half a lattice period, which should lead to a relative suppression of long-range couplings.

To directly determine the fraction of $\ket{b}$ atoms in the bound state, we compare the asymptotic lattice population for a sudden and for an exponential turn-on of the coupling. The results shown in Fig. \ref{FIG:NonMarkov} (D) show that $7.1(2)\%$ of the population are in the evanescent wave, with a total $\ket{b}$ population of $12.7(2)$\%.  The observed fraction of $\ket{b}$ atoms in the bound state (55\%) is close to the expectation \cite{Stewart2017} of $47$\% for the chosen parameters, with the excess possibly stemming from residual non-adiabaticity of the ramping \cite{FootNote2017}.

\paragraph{\bf Conclusion.}
Much of the present work has focused on basic properties arising from the tunability of our Wigner-Weisskopf system, including the formation of bound states below the edge of the mode continuum. On the single-emitter level, this provides a direct analogy to atomic decay near the bandgap of a photonic crystal. We note that, in yet another context, the observed non-Markovian oscillatory dynamics also reproduces predictions for electron photo-detachment from negative ions \cite{Rzazewski1982,Lewenstein2000}.

The optical lattice geometry opens up various additional avenues of inquiry. For emission sufficiently above the continuum edge, these may include novel types of superradiance that depend on the degree of coherence of the lattice population (superfluid or Mott-insulating) \cite{Vega2008,Vega2011} and have no analog in optical systems. Moreover, controlling the longitudinal waveguide level spacing should allow for studies of the transition between the Dicke- and Tavis-Cummings models in quantum optics \cite{Meystre2007, Vega2008} including their modification in the non-Markovian regime. Unlike photons, the emitted atoms can directly interact with each other, which should give rise to additional, nonlinear effects modifying the population dynamics. For negative energies, the bound state lends itself to the realization of lattice models \cite{Vega2008} with modified tunneling and interactions. Superficially, the structure of the bound state resembles that of a lattice polaron \cite{Bruderer2007,Gadway2010} (for which a phononic Lamb-shift has recently been measured \cite{Rentrop2016}), with massive vacuum excitations replacing massless Bogoliubov sound excitations. Rather than reducing transport, the bound state here leads to an enhancement of mobility. The presence of tunneling with a tunable range is, for example, of interest for studies of integrability and thermalization in one-dimensional geometries.

\paragraph{\bf {Acknowledgements.}}
We thank M. G. Cohen for discussions and a critical reading of the manuscript. This work was supported by NSF PHY-1607633. M. S. was supported from a GAANN fellowship by the DoEd. A. P. acknowledges partial support from ESPOL-SENESCYT.

\paragraph{\bf {Author Contributions.}}
D. S., L. K., and M. S. conceived the experiment. L. K. took the measurements with assistance from A. P and J. K. L. K. analyzed the data with contributions by M. S. Numerical simulations were performed by L. K. D. S. supervised the project. The results were discussed and interpreted by all authors. The manuscript was written by L. K. and D. S. with contributions from A. P., J. K. and M. S. 

\paragraph{\bf {\centerline{METHODS}}}

\paragraph{\bf Numerical Simulation.}
We simulate a discrete variation of the problem outlined in \cite{Vega2008} and explored in detail for one 1D-systems in \cite{Stewart2017}, for an array of emitters coupled to a quantized mode structure provided by weak longitudinal harmonic confinement. We start from the Rabi-Hamiltonian (in the RWA) and expand it to couple one or several sites ($\ket{r}$ in the main text) to many different, weakly confined levels ($\ket{b}$ in the main text). This Hamiltonian is (for simplicity only shown for two sites, but readily expanded to n sites)

\begin{equation*}
\hat{H}=\frac{\hbar}{2}
\begin{bmatrix}
2\bar{\delta}_{1}&0&\Omega \gamma_{1,1} &\Omega \gamma_{1,2} & \cdots \\
0&2\bar{\delta}_{2}&\Omega \gamma_{2,1} &\Omega \gamma_{2,2}&  \\
\Omega \gamma_{1,1}& \Omega \gamma_{2,1} &-2\Delta+\omega_z&0 & \\
\Omega \gamma_{1,2}& \Omega \gamma_{2,2} & 0 &-2\Delta+3\omega_z & \\
\vdots & & & & \ddots \\
\end{bmatrix}
\end{equation*}

where $\bar{\delta}_{i}=m\omega_z^2 r_i^2/(2\hbar)$ is a site-dependent detuning (i.e. a site-dependent offset due to the weak harmonic confinement $\omega_z$ experienced by both, the lattice-trapped, and free atoms), and  the $\gamma_{i,j}$ are overlaps between final and initial state wavefunctions (calculated numerically). We use modes up to a fixed frequency ($\omega_{\rm{max}}=2\pi\times5$kHz) and restrict ourselves to $\Delta+2\Omega<\omega_{\rm{max}}$ which yields the insets for Fig. 2 (A) and Fig. 3 (A).

From these simulations we gain additional insight into the effects of the finite system size. As discussed in the main text, the quantized mode structure should act like a true continuum for short enough times where uncertainty should "wash out" the mode structure.  The simulations provide a quantitative test for this.
The simplest comparison is for an isolated emitter in the Markovian limit. We see that for early times, the Markov prediction \cite{Stewart2017} quantitatively agrees with the numerical solution, with a marked deviation (``revival'') observable only at $t=0.5(\omega_z/2\pi)^{-1}$. We note that similar results are also obtained if the continuum is discretized by assuming a periodic-box type potential, where the revival time depends on the length of the box.

While the analytical models \cite{Vega2008,Stewart2017} approach the problem in a strong confinement approximation (where terms of order $(\Delta/\omega_0)^2$ and $(\Omega/\omega_0)^2$ and higher are neglected) our numerical simulation retains all orders. This leads to a slight discrepancy of population transfer that can be noticed between Figs. 2 and 3 (A) and their insets. We furthermore note that our simulation does not reproduce $\delta_L$ for long times.

\paragraph{\bf Background Subtraction.} 

\begin{figure}[h!]
\centering
    \includegraphics[width=0.5\columnwidth]{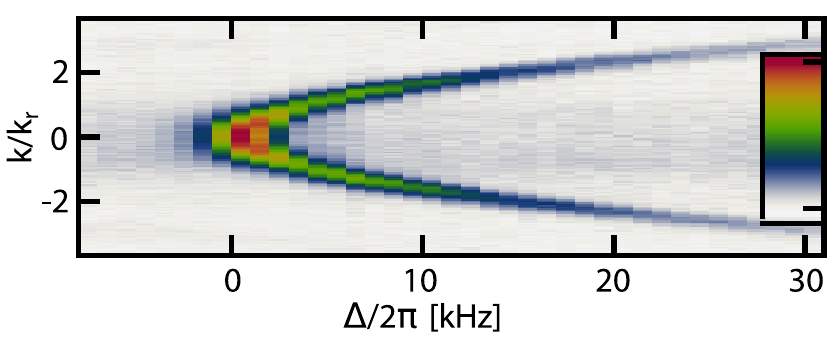}
    \caption{Raw momentum spectrum showing a detuning-independent, diffuse background of $\sim10^3$ atoms.}
    \label{FIG:BGSub}
\end{figure}

The sequence used to thin out the atomic sample leaves roughly $10^3$ atoms in the $\ket{b}$-state before the microwave pulse is applied. This results in a diffuse background in the momentum distributions, as illustrated in Fig. \ref{FIG:BGSub}. We remove this background by subtracting out reference data taken for zero pulse time. The result is Fig.~\ref{FIG:Markov} (C) of the main text.

\paragraph{\bf Momentum calibration.} Our standard momentum calibration relies on Kapitza-Dirac diffraction (KDD) \cite{Gadway2009} from the $z$ lattice. For a more precise determination of emission momenta, we take into account residual propagation in the tubes which slows the atomic motion.  The tubes are created, after ramping up the $z$ lattice, by partial retro-reflection of the Gaussian beams of our optical trap ($1/e^2$ radius of $w=135~\mu$m) \cite{Pertot2009}, which leads to an increase of the optical confinement $\omega_z/2\pi$ from $72(1)~\rm{Hz}$ to $97(1)~\rm{Hz}$  (with gravity along $z$). The tubes are again ramped down within $500~\rm{\mu s}$ after the microwave pulse (together with the $z$ lattice, for band-mapping purposes), followed by a switch off of the optical trap. We numerically simulate the motion of atoms in the tubes, by assuming that the release (with momentum $\pm \bar{k}$) occurs midway through the pulse at the center of the $72~\rm{Hz}$ trap, and by then calculating the trajectories in the time-dependent optical potential until detection after 14.6~ms of time-of-flight. We see that the calibration differs from the KDD results by $-(0.5,1,6)\%$ for pulse durations of $(0.2,0.4,1)$~ms, with negligible differences for shorter pulses. These corrections are included in Fig.~\ref{FIG:Markov} of the main text.

\paragraph{\bf Energy shift data.}

\begin{figure}[h!]
\centering
    \includegraphics[width=0.5\columnwidth]{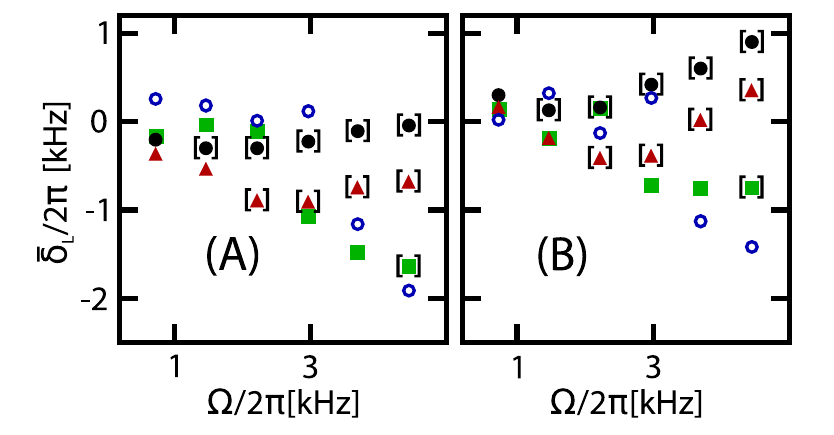}
    \caption{Raw data used to obtain the energy shift as shown in the text. (A) Second moment of k and (B) half separation squared both subtracted by $\Delta/2\pi$. The detunings are $\Delta/2\pi = \left \{ 1000,\,2000,\,4000,\,6000 \right \}~\rm{kHz}$
     for the disks (black), triangles (red), squares (green) and circles (blue) respectively. Points in brackets correspond to the non-Markovian regime $\Omega/\Delta>1$. }
    \label{FIG:LambRaw}
\end{figure}

The main motivation for a precise momentum calibration lies in the smallness of the energetic shift. Another challenge is the blurring of the distribution due to propagation effects for small coupling strengths, i.e. long pulses. We use two measures for the determination of the energy of the emitted wave-packets; the squared separation of the wavepacket-centers (extracted by fitting) and the second moment of the (centered) distribution. The accuracy of the peak separation measure is limited by the fact that it ignores the physical broadening of the momentum distribution at larger coupling strengths and shorter times, while the second moment is sensitive to a blurring of the wavepackets during detection. The data obtained using both methods are shown in Fig.~\ref{FIG:LambRaw}. In the non-Markovian regime $\Omega/\Delta>1$ the peaks become indistinguishable (apex of the parabola in Fig.~\ref{FIG:BGSub}) and a meaningful measure of a shift cannot be extracted with either method.

\normalem


\begin{thebibliography}{37}%
\makeatletter
\providecommand \@ifxundefined [1]{%
 \@ifx{#1\undefined}
}%
\providecommand \@ifnum [1]{%
 \ifnum #1\expandafter \@firstoftwo
 \else \expandafter \@secondoftwo
 \fi
}%
\providecommand \@ifx [1]{%
 \ifx #1\expandafter \@firstoftwo
 \else \expandafter \@secondoftwo
 \fi
}%
\providecommand \natexlab [1]{#1}%
\providecommand \enquote  [1]{``#1''}%
\providecommand \bibnamefont  [1]{#1}%
\providecommand \bibfnamefont [1]{#1}%
\providecommand \citenamefont [1]{#1}%
\providecommand \href@noop [0]{\@secondoftwo}%
\providecommand \href [0]{\begingroup \@sanitize@url \@href}%
\providecommand \@href[1]{\@@startlink{#1}\@@href}%
\providecommand \@@href[1]{\endgroup#1\@@endlink}%
\providecommand \@sanitize@url [0]{\catcode `\\12\catcode `\$12\catcode
  `\&12\catcode `\#12\catcode `\^12\catcode `\_12\catcode `\%12\relax}%
\providecommand \@@startlink[1]{}%
\providecommand \@@endlink[0]{}%
\providecommand \url  [0]{\begingroup\@sanitize@url \@url }%
\providecommand \@url [1]{\endgroup\@href {#1}{\urlprefix }}%
\providecommand \urlprefix  [0]{URL }%
\providecommand \Eprint [0]{\href }%
\providecommand \doibase [0]{http://dx.doi.org/}%
\providecommand \selectlanguage [0]{\@gobble}%
\providecommand \bibinfo  [0]{\@secondoftwo}%
\providecommand \bibfield  [0]{\@secondoftwo}%
\providecommand \translation [1]{[#1]}%
\providecommand \BibitemOpen [0]{}%
\providecommand \bibitemStop [0]{}%
\providecommand \bibitemNoStop [0]{.\EOS\space}%
\providecommand \EOS [0]{\spacefactor3000\relax}%
\providecommand \BibitemShut  [1]{\csname bibitem#1\endcsname}%
\let\auto@bib@innerbib\@empty
\bibitem [{\citenamefont {Weisskopf}\ and\ \citenamefont
  {Wigner}(1930)}]{Weisskopf1930}%
  \BibitemOpen
  \bibfield  {author} {\bibinfo {author} {\bibfnamefont {V.}~\bibnamefont
  {Weisskopf}}\ and\ \bibinfo {author} {\bibfnamefont {E.}~\bibnamefont
  {Wigner}},\ }\bibfield  {title} {\enquote {\bibinfo {title} {{Berechnung der
  nat{\"u}rlichen Linienbreite auf Grund der Diracschen Lichttheorie}},}\
  }\href@noop {} {\bibfield  {journal} {\bibinfo  {journal} {Zeitschrift
  f{\"u}r Physik}\ }\textbf {\bibinfo {volume} {63}},\ \bibinfo {pages}
  {54--73} (\bibinfo {year} {1930})}\BibitemShut {NoStop}%
\bibitem [{\citenamefont {Milonni}(1994)}]{Milonni1994}%
  \BibitemOpen
  \bibfield  {author} {\bibinfo {author} {\bibfnamefont {P.~W.}\ \bibnamefont
  {Milonni}},\ }\href@noop {} {\emph {\bibinfo {title} {{The Quantum Vacuum: An
  Introduction to Quantum Electrodynamics.}}}}\ (\bibinfo  {publisher}
  {Academic Press, Inc.},\ \bibinfo {year} {1994})\BibitemShut {NoStop}%
\bibitem [{\citenamefont {Meystre}\ and\ \citenamefont
  {Sargent~III}(2007)}]{Meystre2007}%
  \BibitemOpen
  \bibfield  {author} {\bibinfo {author} {\bibfnamefont {Pierre}\ \bibnamefont
  {Meystre}}\ and\ \bibinfo {author} {\bibfnamefont {Murray}\ \bibnamefont
  {Sargent~III}},\ }\href@noop {} {\emph {\bibinfo {title} {{Elements of
  Quantum Optics}}}}\ (\bibinfo  {publisher} {Springer Verlag Berlin
  Heidelberg},\ \bibinfo {year} {2007})\BibitemShut {NoStop}%
\bibitem [{\citenamefont {{\'Angel Rivas and Susana F Huelga and Martin B
  Plenio}}(2014)}]{Rivas2014}%
  \BibitemOpen
  \bibfield  {author} {\bibinfo {author} {\bibnamefont {{\'Angel Rivas and
  Susana F Huelga and Martin B Plenio}}},\ }\bibfield  {title} {\enquote
  {\bibinfo {title} {{Quantum non-Markovianity: characterization,
  quantification and detection}},}\ }\href@noop {} {\bibfield  {journal}
  {\bibinfo  {journal} {Reports on Progress in Physics}\ }\textbf {\bibinfo
  {volume} {77}},\ \bibinfo {pages} {094001} (\bibinfo {year}
  {2014})}\BibitemShut {NoStop}%
\bibitem [{\citenamefont {Rotter}\ and\ \citenamefont
  {Bird}(2015)}]{Rotter2015}%
  \BibitemOpen
  \bibfield  {author} {\bibinfo {author} {\bibfnamefont {Ingrid}\ \bibnamefont
  {Rotter}}\ and\ \bibinfo {author} {\bibfnamefont {JP}~\bibnamefont {Bird}},\
  }\bibfield  {title} {\enquote {\bibinfo {title} {{A review of progress in the
  physics of open quantum systems: theory and experiment}},}\ }\href@noop {}
  {\bibfield  {journal} {\bibinfo  {journal} {Reports on Progress in Physics}\
  }\textbf {\bibinfo {volume} {78}},\ \bibinfo {pages} {114001} (\bibinfo
  {year} {2015})}\BibitemShut {NoStop}%
\bibitem [{\citenamefont {Breuer}\ \emph {et~al.}(2016)\citenamefont {Breuer},
  \citenamefont {Laine}, \citenamefont {Piilo},\ and\ \citenamefont
  {Vacchini}}]{Breuer2016}%
  \BibitemOpen
  \bibfield  {author} {\bibinfo {author} {\bibfnamefont {Heinz-Peter}\
  \bibnamefont {Breuer}}, \bibinfo {author} {\bibfnamefont {Elsi-Mari}\
  \bibnamefont {Laine}}, \bibinfo {author} {\bibfnamefont {Jyrki}\ \bibnamefont
  {Piilo}}, \ and\ \bibinfo {author} {\bibfnamefont {Bassano}\ \bibnamefont
  {Vacchini}},\ }\bibfield  {title} {\enquote {\bibinfo {title} {{Colloquium:
  Non-Markovian dynamics in open quantum systems}},}\ }\href@noop {} {\bibfield
   {journal} {\bibinfo  {journal} {Reviews of Modern Physics}\ }\textbf
  {\bibinfo {volume} {88}},\ \bibinfo {pages} {021002} (\bibinfo {year}
  {2016})}\BibitemShut {NoStop}%
\bibitem [{\citenamefont {de~Vega}\ and\ \citenamefont
  {Alonso}(2017)}]{Vega2017}%
  \BibitemOpen
  \bibfield  {author} {\bibinfo {author} {\bibfnamefont {In\'es}\ \bibnamefont
  {de~Vega}}\ and\ \bibinfo {author} {\bibfnamefont {Daniel}\ \bibnamefont
  {Alonso}},\ }\bibfield  {title} {\enquote {\bibinfo {title} {{Dynamics of
  non-Markovian open quantum systems}},}\ }\href@noop {} {\bibfield  {journal}
  {\bibinfo  {journal} {Reviews of Modern Physics}\ }\textbf {\bibinfo {volume}
  {89}},\ \bibinfo {pages} {015001} (\bibinfo {year} {2017})}\BibitemShut
  {NoStop}%
\bibitem [{\citenamefont {Purcell}(1946)}]{Purcell1946}%
  \BibitemOpen
  \bibfield  {author} {\bibinfo {author} {\bibfnamefont {E.~M.}\ \bibnamefont
  {Purcell}},\ }\bibfield  {title} {\enquote {\bibinfo {title} {{Spontaneous
  Emission Probabilities at Radio Frequencies}},}\ }\href@noop {} {\bibfield
  {journal} {\bibinfo  {journal} {Physical Review}\ }\textbf {\bibinfo {volume}
  {69}},\ \bibinfo {pages} {681} (\bibinfo {year} {1946})}\BibitemShut
  {NoStop}%
\bibitem [{\citenamefont {Kleppner}(1981)}]{Kleppner1981}%
  \BibitemOpen
  \bibfield  {author} {\bibinfo {author} {\bibfnamefont {Daniel}\ \bibnamefont
  {Kleppner}},\ }\bibfield  {title} {\enquote {\bibinfo {title} {{Inhibited
  Spontaneous Emission}},}\ }\href@noop {} {\bibfield  {journal} {\bibinfo
  {journal} {Physical Review Letters}\ }\textbf {\bibinfo {volume} {47}},\
  \bibinfo {pages} {233--236} (\bibinfo {year} {1981})}\BibitemShut {NoStop}%
\bibitem [{\citenamefont {Miller}\ \emph {et~al.}(2005)\citenamefont {Miller},
  \citenamefont {Northup}, \citenamefont {Birnbaum}, \citenamefont {Boca},
  \citenamefont {Boozer},\ and\ \citenamefont {Kimble}}]{Miller2005}%
  \BibitemOpen
  \bibfield  {author} {\bibinfo {author} {\bibfnamefont {R.}~\bibnamefont
  {Miller}}, \bibinfo {author} {\bibfnamefont {T.~E.}\ \bibnamefont {Northup}},
  \bibinfo {author} {\bibfnamefont {K.~M.}\ \bibnamefont {Birnbaum}}, \bibinfo
  {author} {\bibfnamefont {A.}~\bibnamefont {Boca}}, \bibinfo {author}
  {\bibfnamefont {A.~D.}\ \bibnamefont {Boozer}}, \ and\ \bibinfo {author}
  {\bibfnamefont {H.~J.}\ \bibnamefont {Kimble}},\ }\bibfield  {title}
  {\enquote {\bibinfo {title} {{Trapped atoms in cavity QED: coupling quantized
  light and matter}},}\ }\href@noop {} {\bibfield  {journal} {\bibinfo
  {journal} {Journal of Physics B: Atomic, Molecular and Optical Physics}\
  }\textbf {\bibinfo {volume} {38}},\ \bibinfo {pages} {S551} (\bibinfo {year}
  {2005})}\BibitemShut {NoStop}%
\bibitem [{\citenamefont {Walther}\ \emph {et~al.}(2006)\citenamefont
  {Walther}, \citenamefont {Varcoe}, \citenamefont {Englert},\ and\
  \citenamefont {Becker}}]{Walther2006}%
  \BibitemOpen
  \bibfield  {author} {\bibinfo {author} {\bibfnamefont {Herbert}\ \bibnamefont
  {Walther}}, \bibinfo {author} {\bibfnamefont {Benjamin~TH}\ \bibnamefont
  {Varcoe}}, \bibinfo {author} {\bibfnamefont {Berthold-Georg}\ \bibnamefont
  {Englert}}, \ and\ \bibinfo {author} {\bibfnamefont {Thomas}\ \bibnamefont
  {Becker}},\ }\bibfield  {title} {\enquote {\bibinfo {title} {{Cavity quantum
  electrodynamics}},}\ }\href@noop {} {\bibfield  {journal} {\bibinfo
  {journal} {Reports on Progress in Physics}\ }\textbf {\bibinfo {volume}
  {69}},\ \bibinfo {pages} {1325} (\bibinfo {year} {2006})}\BibitemShut
  {NoStop}%
\bibitem [{\citenamefont {Haroche}\ and\ \citenamefont
  {Raimond}(2006)}]{Haroche2006}%
  \BibitemOpen
  \bibfield  {author} {\bibinfo {author} {\bibfnamefont {Serge}\ \bibnamefont
  {Haroche}}\ and\ \bibinfo {author} {\bibfnamefont {Jean-Michel}\ \bibnamefont
  {Raimond}},\ }\href@noop {} {\emph {\bibinfo {title} {{Exploring the quantum:
  atoms, cavities, and photons}}}}\ (\bibinfo  {publisher} {Oxford University
  Press},\ \bibinfo {year} {2006})\BibitemShut {NoStop}%
\bibitem [{\citenamefont {John}(1987)}]{John1987}%
  \BibitemOpen
  \bibfield  {author} {\bibinfo {author} {\bibfnamefont {Sajeev}\ \bibnamefont
  {John}},\ }\bibfield  {title} {\enquote {\bibinfo {title} {{Strong
  localization of photons in certain disordered dielectric superlattices}},}\
  }\href@noop {} {\bibfield  {journal} {\bibinfo  {journal} {Physical Review
  Letters}\ }\textbf {\bibinfo {volume} {58}},\ \bibinfo {pages} {2486--2489}
  (\bibinfo {year} {1987})}\BibitemShut {NoStop}%
\bibitem [{\citenamefont {Yablonovitch}(1987)}]{Yablonovich1987}%
  \BibitemOpen
  \bibfield  {author} {\bibinfo {author} {\bibfnamefont {Eli}\ \bibnamefont
  {Yablonovitch}},\ }\bibfield  {title} {\enquote {\bibinfo {title} {{Inhibited
  Spontaneous Emission in Solid-State Physics and Electronics}},}\ }\href@noop
  {} {\bibfield  {journal} {\bibinfo  {journal} {Physical Review Letters}\
  }\textbf {\bibinfo {volume} {58}},\ \bibinfo {pages} {2059--2062} (\bibinfo
  {year} {1987})}\BibitemShut {NoStop}%
\bibitem [{\citenamefont {Lambropoulos}\ \emph {et~al.}(2000)\citenamefont
  {Lambropoulos}, \citenamefont {Nikolopoulos}, \citenamefont {Nielsen},\ and\
  \citenamefont {Bay}}]{Lambropoulos2000}%
  \BibitemOpen
  \bibfield  {author} {\bibinfo {author} {\bibfnamefont {P}~\bibnamefont
  {Lambropoulos}}, \bibinfo {author} {\bibfnamefont {Georgios~M}\ \bibnamefont
  {Nikolopoulos}}, \bibinfo {author} {\bibfnamefont {Torben~R}\ \bibnamefont
  {Nielsen}}, \ and\ \bibinfo {author} {\bibfnamefont {S{\o}ren}\ \bibnamefont
  {Bay}},\ }\bibfield  {title} {\enquote {\bibinfo {title} {{Fundamental
  quantum optics in structured reservoirs}},}\ }\href@noop {} {\bibfield
  {journal} {\bibinfo  {journal} {Reports on Progress in Physics}\ }\textbf
  {\bibinfo {volume} {63}},\ \bibinfo {pages} {455} (\bibinfo {year}
  {2000})}\BibitemShut {NoStop}%
\bibitem [{\citenamefont {Tocci}\ \emph {et~al.}(1996)\citenamefont {Tocci},
  \citenamefont {Scalora}, \citenamefont {Bloemer}, \citenamefont {Dowling},\
  and\ \citenamefont {Bowden}}]{Tocci1996}%
  \BibitemOpen
  \bibfield  {author} {\bibinfo {author} {\bibfnamefont {Michael~D.}\
  \bibnamefont {Tocci}}, \bibinfo {author} {\bibfnamefont {Michael}\
  \bibnamefont {Scalora}}, \bibinfo {author} {\bibfnamefont {Mark~J.}\
  \bibnamefont {Bloemer}}, \bibinfo {author} {\bibfnamefont {Jonathan~P.}\
  \bibnamefont {Dowling}}, \ and\ \bibinfo {author} {\bibfnamefont
  {Charles~M.}\ \bibnamefont {Bowden}},\ }\bibfield  {title} {\enquote
  {\bibinfo {title} {{Measurement of spontaneous-emission enhancement near the
  one-dimensional photonic band edge of semiconductor heterostructures}},}\
  }\href@noop {} {\bibfield  {journal} {\bibinfo  {journal} {Physical Review
  A}\ }\textbf {\bibinfo {volume} {53}},\ \bibinfo {pages} {2799--2803}
  (\bibinfo {year} {1996})}\BibitemShut {NoStop}%
\bibitem [{\citenamefont {Lodahl}\ \emph {et~al.}(2004)\citenamefont {Lodahl},
  \citenamefont {Floris~van Driel}, \citenamefont {Nikolaev}, \citenamefont
  {Irman}, \citenamefont {Overgaag}, \citenamefont {Vanmaekelbergh},\ and\
  \citenamefont {Vos}}]{Lodahl2004}%
  \BibitemOpen
  \bibfield  {author} {\bibinfo {author} {\bibfnamefont {Peter}\ \bibnamefont
  {Lodahl}}, \bibinfo {author} {\bibfnamefont {A.}~\bibnamefont {Floris~van
  Driel}}, \bibinfo {author} {\bibfnamefont {Ivan~S.}\ \bibnamefont
  {Nikolaev}}, \bibinfo {author} {\bibfnamefont {Arie}\ \bibnamefont {Irman}},
  \bibinfo {author} {\bibfnamefont {Karin}\ \bibnamefont {Overgaag}}, \bibinfo
  {author} {\bibfnamefont {Daniel}\ \bibnamefont {Vanmaekelbergh}}, \ and\
  \bibinfo {author} {\bibfnamefont {Willem~L.}\ \bibnamefont {Vos}},\
  }\bibfield  {title} {\enquote {\bibinfo {title} {{Controlling the dynamics of
  spontaneous emission from quantum dots by photonic crystals}},}\ }\href@noop
  {} {\bibfield  {journal} {\bibinfo  {journal} {Nature}\ }\textbf {\bibinfo
  {volume} {430}},\ \bibinfo {pages} {654--657} (\bibinfo {year}
  {2004})}\BibitemShut {NoStop}%
\bibitem [{\citenamefont {Liu}\ \emph {et~al.}(2010)\citenamefont {Liu},
  \citenamefont {Song}, \citenamefont {Wang}, \citenamefont {Bai},
  \citenamefont {Wang}, \citenamefont {Dong}, \citenamefont {Xu},\ and\
  \citenamefont {Han}}]{Liu2010}%
  \BibitemOpen
  \bibfield  {author} {\bibinfo {author} {\bibfnamefont {Qiong}\ \bibnamefont
  {Liu}}, \bibinfo {author} {\bibfnamefont {Hongwei}\ \bibnamefont {Song}},
  \bibinfo {author} {\bibfnamefont {Wei}\ \bibnamefont {Wang}}, \bibinfo
  {author} {\bibfnamefont {Xue}\ \bibnamefont {Bai}}, \bibinfo {author}
  {\bibfnamefont {Yu}~\bibnamefont {Wang}}, \bibinfo {author} {\bibfnamefont
  {Biao}\ \bibnamefont {Dong}}, \bibinfo {author} {\bibfnamefont {Lin}\
  \bibnamefont {Xu}}, \ and\ \bibinfo {author} {\bibfnamefont {Wei}\
  \bibnamefont {Han}},\ }\bibfield  {title} {\enquote {\bibinfo {title}
  {{Observation of Lamb shift and modified spontaneous emission dynamics in the
  YBO 3: Eu 3+ inverse opal}},}\ }\href@noop {} {\bibfield  {journal} {\bibinfo
   {journal} {Optics Letters}\ }\textbf {\bibinfo {volume} {35}},\ \bibinfo
  {pages} {2898--2900} (\bibinfo {year} {2010})}\BibitemShut {NoStop}%
\bibitem [{\citenamefont {Hoeppe}\ \emph {et~al.}(2012)\citenamefont {Hoeppe},
  \citenamefont {Wolff}, \citenamefont {K\"uchenmeister}, \citenamefont
  {Niegemann}, \citenamefont {Drescher}, \citenamefont {Benner},\ and\
  \citenamefont {Busch}}]{Hoeppe2012}%
  \BibitemOpen
  \bibfield  {author} {\bibinfo {author} {\bibfnamefont {Ulrich}\ \bibnamefont
  {Hoeppe}}, \bibinfo {author} {\bibfnamefont {Christian}\ \bibnamefont
  {Wolff}}, \bibinfo {author} {\bibfnamefont {Jens}\ \bibnamefont
  {K\"uchenmeister}}, \bibinfo {author} {\bibfnamefont {Jens}\ \bibnamefont
  {Niegemann}}, \bibinfo {author} {\bibfnamefont {Malte}\ \bibnamefont
  {Drescher}}, \bibinfo {author} {\bibfnamefont {Hartmut}\ \bibnamefont
  {Benner}}, \ and\ \bibinfo {author} {\bibfnamefont {Kurt}\ \bibnamefont
  {Busch}},\ }\bibfield  {title} {\enquote {\bibinfo {title} {{Direct
  Observation of Non-Markovian Radiation Dynamics in 3D Bulk Photonic
  Crystals}},}\ }\href@noop {} {\bibfield  {journal} {\bibinfo  {journal}
  {Physical Review Letters}\ }\textbf {\bibinfo {volume} {108}},\ \bibinfo
  {pages} {043603} (\bibinfo {year} {2012})}\BibitemShut {NoStop}%
\bibitem [{\citenamefont {Bykov}(1975)}]{Bykov1975}%
  \BibitemOpen
  \bibfield  {author} {\bibinfo {author} {\bibfnamefont {Vladimir~P}\
  \bibnamefont {Bykov}},\ }\bibfield  {title} {\enquote {\bibinfo {title}
  {{Spontaneous emission from a medium with a band spectrum}},}\ }\href@noop {}
  {\bibfield  {journal} {\bibinfo  {journal} {Soviet Journal of Quantum
  Electronics}\ }\textbf {\bibinfo {volume} {4}},\ \bibinfo {pages} {861}
  (\bibinfo {year} {1975})}\BibitemShut {NoStop}%
\bibitem [{\citenamefont {John}\ and\ \citenamefont {Wang}(1990)}]{John1990}%
  \BibitemOpen
  \bibfield  {author} {\bibinfo {author} {\bibfnamefont {Sajeev}\ \bibnamefont
  {John}}\ and\ \bibinfo {author} {\bibfnamefont {Jian}\ \bibnamefont {Wang}},\
  }\bibfield  {title} {\enquote {\bibinfo {title} {{Quantum electrodynamics
  near a photonic band gap: Photon bound states and dressed atoms}},}\
  }\href@noop {} {\bibfield  {journal} {\bibinfo  {journal} {Physical Review
  Letters}\ }\textbf {\bibinfo {volume} {64}},\ \bibinfo {pages} {2418--2421}
  (\bibinfo {year} {1990})}\BibitemShut {NoStop}%
\bibitem [{\citenamefont {Liu}\ and\ \citenamefont {Houck}(2017)}]{Liu2017}%
  \BibitemOpen
  \bibfield  {author} {\bibinfo {author} {\bibfnamefont {Yanbing}\ \bibnamefont
  {Liu}}\ and\ \bibinfo {author} {\bibfnamefont {Andrew~A.}\ \bibnamefont
  {Houck}},\ }\bibfield  {title} {\enquote {\bibinfo {title} {{Quantum
  electrodynamics near a photonic bandgap}},}\ }\href@noop {} {\bibfield
  {journal} {\bibinfo  {journal} {Nature Physics}\ }\textbf {\bibinfo {volume}
  {13}},\ \bibinfo {pages} {48--52} (\bibinfo {year} {2017})}\BibitemShut
  {NoStop}%
\bibitem [{\citenamefont {Hood}\ \emph {et~al.}(2016)\citenamefont {Hood},
  \citenamefont {Goban}, \citenamefont {Asenjo-Garcia}, \citenamefont {Lu},
  \citenamefont {Yu}, \citenamefont {Chang},\ and\ \citenamefont
  {Kimble}}]{Hood2016}%
  \BibitemOpen
  \bibfield  {author} {\bibinfo {author} {\bibfnamefont {Jonathan~D.}\
  \bibnamefont {Hood}}, \bibinfo {author} {\bibfnamefont {Akihisa}\
  \bibnamefont {Goban}}, \bibinfo {author} {\bibfnamefont {Ana}\ \bibnamefont
  {Asenjo-Garcia}}, \bibinfo {author} {\bibfnamefont {Mingwu}\ \bibnamefont
  {Lu}}, \bibinfo {author} {\bibfnamefont {Su-Peng}\ \bibnamefont {Yu}},
  \bibinfo {author} {\bibfnamefont {Darrick~E.}\ \bibnamefont {Chang}}, \ and\
  \bibinfo {author} {\bibfnamefont {H.~J.}\ \bibnamefont {Kimble}},\ }\bibfield
   {title} {\enquote {\bibinfo {title} {{Atom–atom interactions around the
  band edge of a photonic crystal waveguide}},}\ }\href@noop {} {\bibfield
  {journal} {\bibinfo  {journal} {Proceedings of the National Academy of
  Sciences}\ }\textbf {\bibinfo {volume} {113}},\ \bibinfo {pages}
  {10507--10512} (\bibinfo {year} {2016})}\BibitemShut {NoStop}%
\bibitem [{\citenamefont {Douglas}\ \emph {et~al.}()\citenamefont {Douglas},
  \citenamefont {Habibian}, \citenamefont {Hung}, \citenamefont {Gorshkov},
  \citenamefont {Kimble},\ and\ \citenamefont {Chang}}]{Douglas2015}%
  \BibitemOpen
  \bibfield  {author} {\bibinfo {author} {\bibfnamefont {J.~S.}\ \bibnamefont
  {Douglas}}, \bibinfo {author} {\bibfnamefont {H}~\bibnamefont {Habibian}},
  \bibinfo {author} {\bibfnamefont {C.~L.}\ \bibnamefont {Hung}}, \bibinfo
  {author} {\bibfnamefont {A.~V.}\ \bibnamefont {Gorshkov}}, \bibinfo {author}
  {\bibfnamefont {H.~J.}\ \bibnamefont {Kimble}}, \ and\ \bibinfo {author}
  {\bibfnamefont {D.~E.}\ \bibnamefont {Chang}},\ }\bibfield  {title} {\enquote
  {\bibinfo {title} {{Quantum many-body models with cold atoms coupled to
  photonic crystals}},}\ }\href@noop {} {\bibfield  {journal} {\bibinfo
  {journal} {Nature Photonics}\ }\textbf {\bibinfo {volume} {9}},\ \bibinfo
  {pages} {326--331}}\BibitemShut {NoStop}%
\bibitem [{\citenamefont {de~Vega}\ \emph {et~al.}(2008)\citenamefont
  {de~Vega}, \citenamefont {Porras},\ and\ \citenamefont
  {Ignacio~Cirac}}]{Vega2008}%
  \BibitemOpen
  \bibfield  {author} {\bibinfo {author} {\bibfnamefont {In\'es}\ \bibnamefont
  {de~Vega}}, \bibinfo {author} {\bibfnamefont {Diego}\ \bibnamefont {Porras}},
  \ and\ \bibinfo {author} {\bibfnamefont {J.}~\bibnamefont {Ignacio~Cirac}},\
  }\bibfield  {title} {\enquote {\bibinfo {title} {{Matter-Wave Emission in
  Optical Lattices: Single Particle and Collective Effects}},}\ }\href@noop {}
  {\bibfield  {journal} {\bibinfo  {journal} {Physical Review Letters}\
  }\textbf {\bibinfo {volume} {101}},\ \bibinfo {pages} {260404} (\bibinfo
  {year} {2008})}\BibitemShut {NoStop}%
\bibitem [{\citenamefont {Navarrete-Benlloch}\ \emph
  {et~al.}(2011{\natexlab{a}})\citenamefont {Navarrete-Benlloch}, \citenamefont
  {de~Vega}, \citenamefont {Porras},\ and\ \citenamefont
  {Cirac}}]{Navarrete2011}%
  \BibitemOpen
  \bibfield  {author} {\bibinfo {author} {\bibfnamefont {Carlos}\ \bibnamefont
  {Navarrete-Benlloch}}, \bibinfo {author} {\bibfnamefont {Inés}\ \bibnamefont
  {de~Vega}}, \bibinfo {author} {\bibfnamefont {Diego}\ \bibnamefont {Porras}},
  \ and\ \bibinfo {author} {\bibfnamefont {J~Ignacio}\ \bibnamefont {Cirac}},\
  }\bibfield  {title} {\enquote {\bibinfo {title} {{Simulating quantum-optical
  phenomena with cold atoms in optical lattices}},}\ }\href@noop {} {\bibfield
  {journal} {\bibinfo  {journal} {New Journal of Physics}\ }\textbf {\bibinfo
  {volume} {13}},\ \bibinfo {pages} {023024} (\bibinfo {year}
  {2011}{\natexlab{a}})}\BibitemShut {NoStop}%
\bibitem [{\citenamefont {Stewart}\ \emph {et~al.}(2017)\citenamefont
  {Stewart}, \citenamefont {Krinner}, \citenamefont {Pazmi\~no},\ and\
  \citenamefont {Schneble}}]{Stewart2017}%
  \BibitemOpen
  \bibfield  {author} {\bibinfo {author} {\bibfnamefont {Michael}\ \bibnamefont
  {Stewart}}, \bibinfo {author} {\bibfnamefont {Ludwig}\ \bibnamefont
  {Krinner}}, \bibinfo {author} {\bibfnamefont {Arturo}\ \bibnamefont
  {Pazmi\~no}}, \ and\ \bibinfo {author} {\bibfnamefont {Dominik}\ \bibnamefont
  {Schneble}},\ }\bibfield  {title} {\enquote {\bibinfo {title} {{Analysis of
  non-Markovian coupling of a lattice-trapped atom to free space}},}\
  }\href@noop {} {\bibfield  {journal} {\bibinfo  {journal} {Physical Review
  A}\ }\textbf {\bibinfo {volume} {95}},\ \bibinfo {pages} {013626} (\bibinfo
  {year} {2017})}\BibitemShut {NoStop}%
\bibitem [{\citenamefont {Pertot}\ \emph {et~al.}(2009)\citenamefont {Pertot},
  \citenamefont {Greif}, \citenamefont {Albert}, \citenamefont {Gadway},\ and\
  \citenamefont {Schneble}}]{Pertot2009}%
  \BibitemOpen
  \bibfield  {author} {\bibinfo {author} {\bibfnamefont {Daniel}\ \bibnamefont
  {Pertot}}, \bibinfo {author} {\bibfnamefont {Daniel}\ \bibnamefont {Greif}},
  \bibinfo {author} {\bibfnamefont {Stephan}\ \bibnamefont {Albert}}, \bibinfo
  {author} {\bibfnamefont {Bryce}\ \bibnamefont {Gadway}}, \ and\ \bibinfo
  {author} {\bibfnamefont {Dominik}\ \bibnamefont {Schneble}},\ }\bibfield
  {title} {\enquote {\bibinfo {title} {{Versatile transporter apparatus for
  experiments with optically trapped Bose-Einstein condensates}},}\
  }\href@noop {} {\bibfield  {journal} {\bibinfo  {journal} {Journal of Physics
  B: Atomic, Molecular and Optical Physics}\ }\textbf {\bibinfo {volume}
  {42}},\ \bibinfo {pages} {215305} (\bibinfo {year} {2009})}\BibitemShut
  {NoStop}%
\bibitem [{\citenamefont {Dicke}(1954)}]{Dicke1954}%
  \BibitemOpen
  \bibfield  {author} {\bibinfo {author} {\bibfnamefont {R.~H.}\ \bibnamefont
  {Dicke}},\ }\bibfield  {title} {\enquote {\bibinfo {title} {{Coherence in
  Spontaneous Radiation Processes}},}\ }\href@noop {} {\bibfield  {journal}
  {\bibinfo  {journal} {Physical Review}\ }\textbf {\bibinfo {volume} {93}},\
  \bibinfo {pages} {99--110} (\bibinfo {year} {1954})}\BibitemShut {NoStop}%
\bibitem [{\citenamefont {$\rm{We}$ noticed an inconsistency within
  \cite{Stewart2017} for the total $\ket{b}$ fraction$\rm{,}$ but this does not
  affect the relative $\ket{b}$ fraction in the~bound state}()}]{FootNote2017}%
  \BibitemOpen
  \bibfield  {author} {\bibinfo {author} {\bibnamefont {$\rm{We}$ noticed an
  inconsistency within \cite{Stewart2017} for the total $\ket{b}$
  fraction$\rm{,}$ but this does not affect the relative $\ket{b}$ fraction in
  the~bound state}},\ }\href@noop {} {}\BibitemShut {NoStop}%
\bibitem [{\citenamefont {Rzazewski}\ \emph {et~al.}(1982)\citenamefont
  {Rzazewski}, \citenamefont {Lewenstein},\ and\ \citenamefont
  {Eberly}}]{Rzazewski1982}%
  \BibitemOpen
  \bibfield  {author} {\bibinfo {author} {\bibfnamefont {K}~\bibnamefont
  {Rzazewski}}, \bibinfo {author} {\bibfnamefont {M}~\bibnamefont
  {Lewenstein}}, \ and\ \bibinfo {author} {\bibfnamefont {JH}~\bibnamefont
  {Eberly}},\ }\bibfield  {title} {\enquote {\bibinfo {title} {{Threshold
  effects in strong-field photodetachment}},}\ }\href@noop {} {\bibfield
  {journal} {\bibinfo  {journal} {Journal of Physics B: Atomic and Molecular
  Physics}\ }\textbf {\bibinfo {volume} {15}},\ \bibinfo {pages} {L661}
  (\bibinfo {year} {1982})}\BibitemShut {NoStop}%
\bibitem [{\citenamefont {Lewenstein}\ and\ \citenamefont
  {Rzazewski}(2000)}]{Lewenstein2000}%
  \BibitemOpen
  \bibfield  {author} {\bibinfo {author} {\bibfnamefont {M.}~\bibnamefont
  {Lewenstein}}\ and\ \bibinfo {author} {\bibfnamefont {K.}~\bibnamefont
  {Rzazewski}},\ }\bibfield  {title} {\enquote {\bibinfo {title} {{Quantum
  anti-Zeno effect}},}\ }\href@noop {} {\bibfield  {journal} {\bibinfo
  {journal} {Physical Review A}\ }\textbf {\bibinfo {volume} {61}},\ \bibinfo
  {pages} {022105} (\bibinfo {year} {2000})}\BibitemShut {NoStop}%
\bibitem [{\citenamefont {Navarrete-Benlloch}\ \emph
  {et~al.}(2011{\natexlab{b}})\citenamefont {Navarrete-Benlloch}, \citenamefont
  {de~Vega}, \citenamefont {Porras},\ and\ \citenamefont {Cirac}}]{Vega2011}%
  \BibitemOpen
  \bibfield  {author} {\bibinfo {author} {\bibfnamefont {Carlos}\ \bibnamefont
  {Navarrete-Benlloch}}, \bibinfo {author} {\bibfnamefont {Inés}\ \bibnamefont
  {de~Vega}}, \bibinfo {author} {\bibfnamefont {Diego}\ \bibnamefont {Porras}},
  \ and\ \bibinfo {author} {\bibfnamefont {J~Ignacio}\ \bibnamefont {Cirac}},\
  }\bibfield  {title} {\enquote {\bibinfo {title} {{Simulating quantum-optical
  phenomena with cold atoms in optical lattices}},}\ }\href@noop {} {\bibfield
  {journal} {\bibinfo  {journal} {New Journal of Physics}\ }\textbf {\bibinfo
  {volume} {13}},\ \bibinfo {pages} {023024} (\bibinfo {year}
  {2011}{\natexlab{b}})}\BibitemShut {NoStop}%
\bibitem [{\citenamefont {Bruderer}\ \emph {et~al.}(2007)\citenamefont
  {Bruderer}, \citenamefont {Klein}, \citenamefont {Clark},\ and\ \citenamefont
  {Jaksch}}]{Bruderer2007}%
  \BibitemOpen
  \bibfield  {author} {\bibinfo {author} {\bibfnamefont {Martin}\ \bibnamefont
  {Bruderer}}, \bibinfo {author} {\bibfnamefont {Alexander}\ \bibnamefont
  {Klein}}, \bibinfo {author} {\bibfnamefont {Stephen~R.}\ \bibnamefont
  {Clark}}, \ and\ \bibinfo {author} {\bibfnamefont {Dieter}\ \bibnamefont
  {Jaksch}},\ }\bibfield  {title} {\enquote {\bibinfo {title} {{Polaron physics
  in optical lattices}},}\ }\href@noop {} {\bibfield  {journal} {\bibinfo
  {journal} {Physical Review A}\ }\textbf {\bibinfo {volume} {76}},\ \bibinfo
  {pages} {011605} (\bibinfo {year} {2007})}\BibitemShut {NoStop}%
\bibitem [{\citenamefont {Gadway}\ \emph {et~al.}(2010)\citenamefont {Gadway},
  \citenamefont {Pertot}, \citenamefont {Reimann},\ and\ \citenamefont
  {Schneble}}]{Gadway2010}%
  \BibitemOpen
  \bibfield  {author} {\bibinfo {author} {\bibfnamefont {Bryce}\ \bibnamefont
  {Gadway}}, \bibinfo {author} {\bibfnamefont {Daniel}\ \bibnamefont {Pertot}},
  \bibinfo {author} {\bibfnamefont {Ren\'e}\ \bibnamefont {Reimann}}, \ and\
  \bibinfo {author} {\bibfnamefont {Dominik}\ \bibnamefont {Schneble}},\
  }\bibfield  {title} {\enquote {\bibinfo {title} {{Superfluidity of
  Interacting Bosonic Mixtures in Optical Lattices}},}\ }\href@noop {}
  {\bibfield  {journal} {\bibinfo  {journal} {Physical Review Letters}\
  }\textbf {\bibinfo {volume} {105}},\ \bibinfo {pages} {045303} (\bibinfo
  {year} {2010})}\BibitemShut {NoStop}%
\bibitem [{\citenamefont {Rentrop}\ \emph {et~al.}(2016)\citenamefont
  {Rentrop}, \citenamefont {Trautmann}, \citenamefont {Olivares}, \citenamefont
  {Jendrzejewski}, \citenamefont {Komnik},\ and\ \citenamefont
  {Oberthaler}}]{Rentrop2016}%
  \BibitemOpen
  \bibfield  {author} {\bibinfo {author} {\bibfnamefont {T.}~\bibnamefont
  {Rentrop}}, \bibinfo {author} {\bibfnamefont {A.}~\bibnamefont {Trautmann}},
  \bibinfo {author} {\bibfnamefont {F.~A.}\ \bibnamefont {Olivares}}, \bibinfo
  {author} {\bibfnamefont {F.}~\bibnamefont {Jendrzejewski}}, \bibinfo {author}
  {\bibfnamefont {A.}~\bibnamefont {Komnik}}, \ and\ \bibinfo {author}
  {\bibfnamefont {M.~K.}\ \bibnamefont {Oberthaler}},\ }\bibfield  {title}
  {\enquote {\bibinfo {title} {{Observation of the Phononic Lamb Shift with a
  Synthetic Vacuum}},}\ }\href@noop {} {\bibfield  {journal} {\bibinfo
  {journal} {Physical Review X}\ }\textbf {\bibinfo {volume} {6}},\ \bibinfo
  {pages} {041041} (\bibinfo {year} {2016})}\BibitemShut {NoStop}%
\bibitem [{\citenamefont {Gadway}\ \emph {et~al.}(2009)\citenamefont {Gadway},
  \citenamefont {Pertot}, \citenamefont {Reimann}, \citenamefont {Cohen},\ and\
  \citenamefont {Schneble}}]{Gadway2009}%
  \BibitemOpen
  \bibfield  {author} {\bibinfo {author} {\bibfnamefont {Bryce}\ \bibnamefont
  {Gadway}}, \bibinfo {author} {\bibfnamefont {Daniel}\ \bibnamefont {Pertot}},
  \bibinfo {author} {\bibfnamefont {Ren{\'e}}\ \bibnamefont {Reimann}},
  \bibinfo {author} {\bibfnamefont {Martin~G}\ \bibnamefont {Cohen}}, \ and\
  \bibinfo {author} {\bibfnamefont {Dominik}\ \bibnamefont {Schneble}},\
  }\bibfield  {title} {\enquote {\bibinfo {title} {{Analysis of Kapitza-Dirac
  diffraction patterns beyond the Raman-Nath regime}},}\ }\href@noop {}
  {\bibfield  {journal} {\bibinfo  {journal} {Optics Express}\ }\textbf
  {\bibinfo {volume} {17}},\ \bibinfo {pages} {19173--19180} (\bibinfo {year}
  {2009})}\BibitemShut {NoStop}%
\end{thebibliography}
\end{document}